\documentstyle[aps,amsbsy,epsfig,twocolumn]{revtex}
\begin{document}

\def \virg{\;\;,}
\def \point{\;\;.}
\def \up{\uparrow}
\def \down{\downarrow}
\def \ve#1{\mbox{\boldmath $#1$}}
\def \eps{\varepsilon}
\def \omn#1{\ve{\Omega}_{#1}}
\newcommand{\rs}{{\bf r}s}
\newcommand{\k}{{\bf k}}
\newcommand{\be}{\begin{equation}}
\newcommand{\ee}{\end{equation}}
\newcommand{\r}{{\bf r}}
\newcommand{\Q}{{\bf Q}}
\newcommand{\p}{{\bf p}}
\newcommand{\vv}{{\bf v}}
\newcommand{\q}{{\bf q}}
\newcommand{\rp}{{{\bf r}'}}
\newcommand{\rpp}{{{\bf r}''}}
\newcommand{\rsp}{{\bf r}'s}
\newcommand{\ru}{{{\bf r} \up}}
\newcommand{\rd}{{{\bf r} \down}}
\newcommand{\tr}{{\rm tr}}
\newcommand{\av}[2]{{\left\langle #1 \right\rangle}_{#2} }
\newcommand{\bal}{{\rm bal.}}
\newcommand{\cN}{{\cal N}}
\newcommand{\cZ}{{\cal Z}}
\newcommand{\diff}{{\rm diff.}}
\newcommand{\dN}{\langle\delta N^2(\phi)\rangle}
\newcommand{\Ec}{E_c}
\newcommand{\Ef}{E_F}
\newcommand{\ens}{{\rm ens}}
\newcommand{\ep}{\epsilon}

\newcommand{\bep}{\boldsymbol\epsilon}
\newcommand{\bra}[1]{\left\langle#1\right|}
\newcommand{\ket}[1]{\left|#1\right\rangle}

\draft

\title{Phase coherence times in the multiple scattering of photons by 
cold
atoms} 
\author{E.~Akkermans$^1$, Ch.~Miniatura$^2$ and C.A.~M\"uller$^2$\\
$^1$ Department of Physics, Technion, 32000 Haifa, Israel \\ 
$^2$ Laboratoire Ondes et D\'esordre (FRE
2302 CNRS), 1361 route des Lucioles \\ F-06560 Valbonne, France\\}

\maketitle

\begin{abstract}
We present an analysis of the dephasing present 
in the multiple scattering of photons by atoms with a quantum internal
structure. 
The corresponding phase coherence  
times $\tau_{\phi}$ are obtained as a function of the Zeeman 
degeneracy of the atomic dipole transition and the polarization state
of the photons.  
These results allow for an explanation of the recent
experiments on coherent 
backscattering of photons from a gas of cold rubidium atoms where the 
height of the backscattering cone
depends on the atomic internal degrees of freedom 
coupled to the polarization of the photons. Some consequences of 
these 
results are presented, and analogies with the case of electronic
systems are highlighted.  
\end{abstract}
 

\def\real{{\rm I\kern-.2em R}}
\def\complex{\kern.1em{\raise.47ex\hbox{
	    $\scriptscriptstyle |$}}\kern-.40em{\rm C}}
\def\integer{{\rm Z\kern-.32em Z}}
\def\pinteger{{\rm I\kern-.15em N}}

Multiple scattering of waves in complex media accounts for
a large number of coherent effects which are often presented 
under the shortcut of mesoscopic physics. These coherent effects 
result from interferences between 
the amplitudes associated to 
multiple 
scattering paths of the wave. In the weak scattering limit for which 
$k 
l_{e} \gg 1$, where $k$ is the wave-vector and $l_{e}$ the elastic 
mean 
free path (the average distance between successive collisions), the 
relevant physical quantities can be obtained from the probability 
$P(\r,\r',t)$ of quantum diffusion defined as the disorder averaged 
probability to propagate a wave packet between the points $\bf r$ and
$\bf r'$ in a time $t$ \cite{am,Chakraverty86}. It allows to describe 
either weak localization effects
in metals \cite{am,bergman}, spectral quantities or
the coherent albedo and dynamical effects in multiple scattering of 
light 
by classical scatterers \cite{am}.

Interference effects are very sensitive to dephasing. Roughly 
speaking, a dephasing may originate either from an external field 
\cite{bergman,sharvin} or 
from additional 
degrees of freedom which affect in different ways each of the 
interfering
amplitudes. An example is provided by the 
spin-flip scattering 
in metals where the spin of the electron rotates due to scattering by
magnetic impurities \cite{bergman}.

In the presence of dephasing, the probability of 
quantum diffusion can be written as $ P(\r, \r', t) \left\langle e^{i 
\phi
(t)} \right\rangle$,  
where the random variable $\phi(t)$ is the relative phase of the two 
interfering paths. Its distribution depends on 
the origin of the dephasing and we denote by $\langle ...\rangle$ 
the average over this distribution. In most cases we have 
$\left\langle 
e^{i   \phi (t)} \right\rangle \simeq e^{- t/ \tau_\phi}$ at least for 
long enough times $t$. The characteristic time $\tau_\phi$ is the 
phase coherence or dephasing time.
An exponential decrease of the probability of quantum diffusion 
does not necessarily describe a dephasing process. For 
instance, the intensity of an electromagnetic wave which propagates 
in an
absorbing medium decreases exponentially. 
But this is not a dephasing process since it affects 
equally both the coherent and incoherent contributions by a decrease 
of 
the overall intensity. We propose to speak of dephasing only when the 
coherent
part is reduced with respect to the incoherent part. 

Recently, the dephasing induced by the Coulomb interactions in metals 
has been reconsidered in detail in view of the recent experiments on 
the dephasing time $\tau_{\phi} (T)$ of the electrons at the Fermi 
level where an unexpected saturation at low temperature has been 
observed \cite{webb}.

The propagation of photons in complex systems addresses similar 
questions and provides new sources of dephasing like the motion of 
the 
scatterers \cite{SO}. It also provides access to the type of 
scattering
experienced by the  
photons (Rayleigh, Mie, resonant, Raman etc.) so that we 
have at our disposal both a theoretical description of the elementary 
scattering events and its generalization to multiple scattering. 

Dephasing and decoherence in multiple scattering are 
sometimes perceived as a nuisance which prevents the observation of 
full interference effects. Instead, when the dephasing is well 
controlled, it should better be viewed as a 
unique tool in order to obtain new insights on the system that are 
usually out of reach in the very dilute and single scattering limits. 
A beautiful implementation 
of this has been demonstrated in the so-called diffusive wave 
spectroscopy \cite{dws}. There, the controlled dephasing 
between 
multiple scattering trajectories of the photons resulting from the
dynamics of the  
scatterers allows to probe this dynamics on time scales unreachable 
otherwise. 

Recently, the problem of multiple scattering of photons in a gas of 
cold atoms has been investigated in detail and coherent 
backscattering has been observed \cite{cbsRb,barrat}. It has been shown that 
the 
existence of internal atomic degrees of freedom modifies 
significantly 
the coherent backscattering \cite{Mueller01}.  
The purpose of this 
letter is to present an analysis of the dephasing induced by the 
internal atomic degrees of freedom and to derive corresponding
expressions for the  
phase coherent times as a function of the polarization state of the 
photons and of the Zeeman degeneracy.  We believe that the present 
analysis 
provides relevant spectroscopic tools in order to probe the dynamics 
of
cold atoms.

We describe a gas of $N$ atoms as 
two-level systems of characteristic transition frequency $\omega_{0}$ 
\cite{Mueller01}. 
The ground state defines the zero of energy and has total angular 
momentum 
$J$. The excited state has a total angular momentum $J_{e}$ and a 
natural width $\Gamma$ due to coupling to the vacuum 
fluctuations. We shall assume, moreover, that the velocity $v$ of the 
atoms is small compared to $\Gamma / k$ ($k$ is the light 
wave-vector) but large compared to $\hbar k / M$ ($M$ 
being the mass of the atom), so that it is possible to neglect the 
Doppler 
and recoil effects. The external degrees of freedom of the atoms are  
therefore the classical assigned positions $\r_{\alpha} \ \ (\alpha 
=1,\ldots,N$) uncorrelated with one another. The corresponding 
one-atom 
Hamiltonian 
\begin{equation}
H_{at} = \omega_{0} \sum_{m_{e} = - J_{e}}^{J_{e}} |J_{e} m_{e} 
\rangle \langle J_{e} m_{e}|
\end{equation}
describes the internal quantum degrees of freedom, in units where 
$\hbar = c =1$. No magnetic field is supposed to be present, so that 
the two levels are respectively $(2 J +1)$ and $(2 J_{e} + 1)$-fold
degenerate.  

The atom-photon interaction is described within 
the dipole approximation, namely using the Hamiltonian
\be
V=-\sum_{\alpha=1}^{N}{\bf D}_\alpha\cdot{\bf E}(\r_\alpha) 
\ee
where ${\bf D}_{\alpha}$ is the atomic dipole operator and ${\bf E}
(\r)$ is the quantized electric field operator  
(quantization 
volume ${\cal V}$) 
evaluated at the center of mass $\r_{\alpha}$ of each atom. 
The natural width of
the excited atomic state is $\Gamma= d^2\omega_0^3/3\pi\epsilon_0$ 
where $d = \langle J_{e} || {\bf D} || J \rangle / \sqrt{2 
J_{e} +1 }$ in standard notations.   We define the 
dimensionless dipole operator 
${\bf d} = {\bf  D}/d$
with non vanishing matrix 
elements only between the two states $J$ and $J_{e}$. 
The elastic scattering process between the two states $|\k \bep,Jm 
\rangle$ and $|\k' \bep',Jm' \rangle$, where $|\k\bep\rangle$ is a 
one-photon Fock state of the free transverse electromagnetic field in 
the mode
$\k$  
of polarization $\bep$, is described by the single scattering
transition amplitude 
$t_{ij} (m,m',\omega) = t(\omega) \langle J m'|d_{i} d_{j}| J m 
\rangle$. 
Here,  the resonant scattering amplitude is given by 
\be
t(\omega) = {3 \over 2 \pi \rho_{0} (\omega)} {\Gamma /2  \over 
\delta + i \Gamma / 2}
\ee
where $\delta = \omega - \omega_{0}$ is the detuning of the probe 
light 
from the atomic resonance and $\rho_0(\omega)={\cal V}\omega^2/2\pi^2$ is 
the free photon spectral
density.  
The amplitude $t_{ij}$ is a $3 \times 3$-matrix which connects the 
incoming and outgoing polarizations of the scattering photon. It can 
be 
decomposed into the sum of a scalar, an antisymmetric and a traceless 
symmetric part. The scalar part is the only one which remains in the 
case of the Rayleigh scattering on a classical dipole ($J=0,J_e=1$).  

In order to characterize the multiple scattering 
of a photon of frequency $\omega$, we first calculate its average 
propagator 
$ \overline{G} (\omega)$. The averaging is over the 
uncorrelated positions $\r_{\alpha}$ of the atoms and over the 
magnetic
quantum numbers $m_\alpha$ of the atoms. The first, standard  average 
restores the translation invariance. The internal average, a trace 
with a
scalar density matrix $\rho$ 
assuming that the atoms are prepared independently and equally in 
their ground 
states, restores rotational invariance. 
The calculation of the average propagator reduces to that  
of the scalar self-energy $\Sigma_{ij}
(\omega)=\Sigma(\omega) \delta_{ij}$. For a dilute enough atomic gas, 
it can be calculated within the self-consistent Born approximation
\cite{am}
which neglects all possible interference originating from photon 
exchange between distinct atoms:
$\Sigma (\omega) = N M_{J}\,  t(\omega)$  
where $M_{J} = \frac{1}{3}(2 J_{e} +1)/ (2J+1)$ such that $M_{0} =1$
\cite{Mueller01}.  
The real and the imaginary parts of the self-energy give 
respectively the photon frequency shift and the elastic scattering 
mean free time $\tau_{e}=-[2\mbox{Im} \Sigma (\omega)]^{-1}$. The 
elastic mean 
free path $l_{e} = \tau_{e}$ (remember $c=1$) 
measures the average distance between two 
successive collisions of the photons. It is worth 
emphasizing that $l_{e}$, up to the scalar factor $M_{J}$, is 
independent of
the internal quantum 
structure of the atoms. Indeed, the optical theorem 
assures that the total cross section $\sigma=1/nl_e$ 
equals the imaginary part of the diagonal matrix element of the 
collision amplitude $t_{ij} (\omega)$ which depends on the scalar 
component of this tensor only.

To go further, we need to calculate the (time 
integrated) average 
probability $P(\r_{0}, \r)$ for a photon to propagate from a point 
source $\r_{0}$ to a point $\r$. 
We will calculate its Fourier transform
\be 
P_{\k\k'}
(\q) \propto 
\overline{\bra{\k-{\q\over2}}G^A\ket{\k'-{\q\over2}}
\bra{\k'+{\q\over2}}G^R\ket{\k+{\q\over2}}}
\ee
rewritten as 
$
P_{\k\k'} (\q) \propto 
\overline{G}{}^A \overline{G}{}^R +\overline{G}{}^A 
\overline{G}{}^R \,
\Gamma_{\k\k'}(\q)\, 
\overline{G}{}^A \overline{G}{}^R$.
The first term is the Drude-Boltzmann contribution. In the weak 
scattering limit $k l_{e} \gg 1$, the second term
 contains two contributions: the incoherent 
diffuson $\Gamma_{d}$ for which the phase averages to zero, and the 
coherent 
cooperon $\Gamma_{c}$ which still retains an overall phase. 
In the limit of dilute atomic gases, it is possible to write for both 
the 
diffuson and the cooperon an integral equation 
which generalizes the well-known scalar calculation \cite{am} and 
takes into account the change of the photon polarization as a result 
of the existence of internal degrees of freedom of the atoms 
\cite{grospapier}.

In order to describe multiple scattering, 
we need first to determine the atomic 
intensity vertex $U$ (the differential cross section), the average 
square of
the 
matrix element 
$\langle \k' \bep' | T(\omega)| \k \bep 
\rangle = (\overline{ \bep}'\cdot t(\omega)\cdot \bep) \,  e^{i(\k - 
\k')\cdot 
\r}$ of the transition operator. Averaging over the position and the 
internal quantum numbers yields
\cite{Mueller01}
\begin{eqnarray}
U &=& N \mbox{Tr}\left[\rho\, 
\left(\overline{ \bep}_{4}.t(\omega). \bep_{3} \right) \left( 
\overline{ \bep}_{2}.t(\omega). \bep_{1} \right) \right] 
\delta_{\k_{1}+ \k_{3},\k_{2}+ \k_{4}}
\nonumber \\
&=& M_{J} |t(\omega)|^{2} 
{ \ep}_{1i} \overline{{ \ep}}_{2j} { \ep}_{3k} 
\overline{{ \ep}}_{4l} \, 
I_{il,jk} \delta_{\k_{1}+ \k_{3},\k_{2}+ \k_{4}}. 
\end{eqnarray}
The vertex $U$ describes the scattering of two incoming 
photons $(\k_{1}, { \bep}_{1})$ and $(\k_{3}, { \bep}_{3})$ into the 
two outgoing states $(\k_{2}, { \bep}_{2})$ and $(\k_{4}, { 
\bep}_{4})$. 
The rank-four tensor 
$
I_{il,jk} = M_{J}^{-1} \mbox{Tr} \left[ \rho \, d_l d_k d_j 
d_i\right] 
$
can be decomposed into its three irreducible 
components \cite{grospapier}
\be
I_{il,jk} = \sum_{K=0}^{2} \lambda_{K} \, T^{(K)}_{il,jk}
\label{intvertdiff}
\ee
where the eigenvalues 
\be
\lambda_{K} =  3 (2 J_{e} +1) \left\{ \begin{array}{clcr}
1 & 1 & K \\
J_{e} & J_{e} & J 
\end{array}
\right\}^{2}
\ee
are given in terms of $6j$-Wigner symbols.  The three basic tensors  
\begin{eqnarray}
T^{(0)}_{il,jk} &=& {1\over3}\delta_{il}\delta_{jk} \label{T0}\\
T^{(1)}_{il,jk} &=&
{1\over2}\left[\delta_{ij}\delta_{kl}-\delta_{ik}\delta_{jl}\right] 
\label{T1}\\
T^{(2)}_{il,jk} &=&
{1\over2}\left[\delta_{ij}\delta_{kl}+\delta_{ik}\delta_{jl}\right]
-{1\over3}\delta_{il}\delta_{jk} \label{T2}
\end{eqnarray}
project on the scalar, 
antisymmetric and symmetric traceless components, respectively,  of 
the field intensity matrix $E_iE_j$.      
Note that energy conservation imposes $\lambda_0=1$ for all $J,J_e$. 
For the pure dipole scatterer $J=0,J_e=1$,  
the scattering tensor is proportional to the identity, implying that 
all eigenvalues are $\lambda_K=1$. 

Then, the propagation of the intensity 
between successive scattering events is described by the product of 
the average photon propagators, namely 
\be
G_{il,jk} (\q) = {3 \over 8 \pi l_{e}} \int d^{3} r {e^{-r/l_{e}} \over 
r^{2}} 
\Delta_{ij} \Delta_{kl} e^{i \q.\r}
\ee
where $\Delta_{ij} = \delta_{ij} - {\hat r}_{i}{\hat r}_{j}$ 
describes the
transverse projection of the polarization 
vector. In the
diffusive limit $q l_{e} \ll 1$, the intensity propagator 
$G_{il,jk}$ can also be decomposed using the basic tensors 
(\ref{T0}-\ref{T2}) and we recover the 
eigenvalues $b_{0} = 1$, $b_{1} ={1\over2}$ and $b_{2} ={7\over 10}$
as for the classical  dipolar case \cite{polarisation}.

The summation of the geometric series built from the elementary 
scattering event, i.e.\ the tensorial product of $I_{il,jk} $ and 
$G_{il,jk}$, yields in the diffusive limit $q l_{e} \ll 1$ for the 
diffuson 
\be
\Gamma_{d}(\q) = {3 \over  4 \pi \rho_{0}  l_{e} \tau_{\rm tr}}  
\sum_{K=0}^{2} \Lambda_{K}
(q) \,  { \ep}_{1i} \overline{{ \ep}}_{2j} { \ep}_{3k} 
\overline{{ \ep}}_{4l} \,   T^{(K)}_{il,jk} 
\ee
where
the propagators of the three 
eigenmodes are 
\be
\Lambda_{K } (q)
\label{Lambda} = {1\over b_K}{1\over Dq^2 + \tau_{d}^{-1}(K)}. 
\ee
Here, $D= l^2_{e} / 3\tau_{\rm tr}$ is the diffusion constant, 
involving the
resonant transport time scale $\tau_{\rm tr}=\l_e+\Gamma^{-1}$
\cite{grospapier,vanAlbada91}.   
The polarization relaxation times 
\be
\tau_{d}(K) = {b_K\lambda_K \over 1 - b_K \lambda_K }\tau_{\rm tr}
\label{taudep}
\ee
describe the depolarization of the initial 
light beam 
and are related to the 
depolarization factors 
calculated in the classical Rayleigh case \cite{awmm}. Among the 
three modes,
we have   
$ \tau_{d}^{-1}(0)=0$, and the corresponding scalar mode 
$\Lambda_{0}(q)$ diverges for small $q$. It is the singlet Goldstone 
mode associated  
with the local conservation of the number of photons. The 
antisymmetric mode $K=1$ is analogous to the triplet mode obtained in 
the spin-orbit scattering in electronic systems \cite{bergman}. 

A finite time $\tau_{d}$ 
leads to an exponential attenuation of incoherent diffusion 
contribution. In order to identify unambiguously a dephasing time, we have to
consider the cooperon contribution $\Gamma_{c}$. It 
accounts for the interference of amplitudes associated to two 
time-reversed 
multiple scattering paths of the photons and can be calculated along 
the same lines. 
The atomic intensity vertex has a form similar to 
(\ref{intvertdiff}), with the new eigenvalues defined in terms of 
$9j$-symbols, 
\be
\chi_{K} = 3(2J_e+1)
\left\{ \begin{array}{ccc}
1 & J_e & J \\
1 & J & J_{e}\\
K & 1 & 1 
\end{array}
\right\} 
\ee 
The cooperon contribution reads 
\be
\Gamma_{c}(\q_c) = {3  \over  4 \pi \rho_{0} l_{e} \tau_{\rm tr}}  
\sum_{K} X_{K}
(q_c) \,  { \ep}_{1i} \overline{{ \ep}}_{2j}  
\overline{{ \ep}}_{4k} { \ep}_{3l} \,   T^{(K)}_{il,jk}
\ee
where $\q_c=\k+\k'$ is the total momentum and the polarization 
vectors ${\bep}_{3}$ and $\overline{{ \bep}}_{4}$ have been 
exchanged. The propagators of the three eigenmodes are given by 
\be
X_{K}(q) = {\chi_{K} \over D q^{2} + 
\tau_{d}^{-1}(K) + {\tau_{\phi}^{-1}(K)}}
\ee
The overall depolarization described by the times $\tau_{d}(K)$ 
affects the cooperon as well. But the cooperon involves an 
additional contribution, the dephasing time 
\be
\tau_{\phi}(K) = {\lambda_K\chi_K \over \lambda_{K} - \chi_{K}} \, 
\tau_{\rm tr}  
\label{tauphi}
\ee
This expression constitutes our main result. 
The limit of the classical 
Rayleigh scattering $J=0$ corresponds to $\lambda_{K} = 
\chi_{K} =1 $ and thus to the absence of dephasing ($\tau_{\phi}(0)=\infty $). 
Of particular interest is the value of $\tau_{\phi}(0)$ associated to 
the
intensity of the field. Explicitly, we find 
\be
\frac{\tau_\phi(0)}{\tau_{\mathrm tr}} = 
\left\{ \begin{array}{ll}
	(J(2J+3))^{-1},& \quad  J_e=J+1 \\
	J^2+J-1, & \quad J_e=J\\
	(2J^2+J-1)^{-1},& \quad J_e=J-1 
	\end{array}
\right.
\ee
An absence of dephasing only occurs for the 
classical
dipole $J=0$ and in the semi-classical limit $J = J_e \to \infty$. 

 In experiments on coherent backscattering (CBS) of light 
by atoms without an internal
degeneracy ($J=0$), the optimal CBS enhancement factor is found in 
the 
polarization channel of preserved helicity \cite{Bidel02}. 
The present analysis allows to explain the unexpected experimental
observation that the CBS enhancement factor for cold rubidium
atoms ($J=3,J_e=4$) in the channel of preserved helicity  is lower 
than in the channel of flipped helicity \cite{cbsRb}. Indeed, 
following
\cite{awmm}, we define a CBS contrast function ${\cal
C}(n)=\Gamma_c^{(n)}(0)/\Gamma_d^{(n)}(0)$ as the ratio of cooperon and 
diffuson 
contributions at the $n$th scattering order. For large $n$, only the 
largest
eigenvalues $b_0=\lambda_0=1$ of the diffuson mode expansion will
contribute for both polarization channels. For the channel of flipped
helicity  ($\perp$), the contrast function behaves as ${\cal 
C}_\perp(n)\sim \frac{30}{7}
(b_2\chi_2)^n$ and decreases exponentially.  
In the channel of preserved helicity ($\parallel$), all irreducible
modes contribute {\em a priori}: 
${\cal C}_\parallel(n) \sim \chi_0^n-3 (b_1\chi_1)^n + {5\over7}
(b_2\chi_2)^n$.  
In the case of classical dipole scatterers such that 
$\chi_K=1$, the parallel contrast stays optimal, ${\cal 
C}_\parallel(n) \sim
1$. But in the case $J=3,J=4$, the product $b_2\chi_2={19 \over 
40}$ is much larger than $\chi_0={1\over28}$ and therefore
dominates. The contrast in the perpendicular channel then is higher 
than in the
parallel channel, ${\cal C}_\perp/ {\cal C}_\parallel \sim 6$, 
independently of
the scattering order $n$. 
 
In terms of the above definitions, 
$b_K\chi_K = \left[1+\tau_{\mathrm{tr}}/\tau_d(K) +
	\tau_{\mathrm{tr}}/\tau_\phi(K)\right]^{-1}$. 
We can therefore conclude that the loss of CBS contrast in 
perpendicular
polarization channels is not a dephasing process since it persists 
even for
$J=0$ where 
$\tau_\phi^{-1}(K)=0$, but still $\tau_d^{-1}(2)>0$. 
Since the cooperon and diffuson contributions probe  
different field correlators, it should rather be regarded as an 
effect of
polarization decorrelation. On the contrary, an atomic internal 
degeneracy
affects the contrast in channels of parallel polarizations and 
therefore
appears as a particularly neat realization of a microscopic dephasing 
mechanism.     

In summary, we have identified dephasing times in the weak
localization of light by cold atoms. They depend in a remarkably
simple manner on the internal atomic degeneracy and the field 
polarization
mode. It is of interest to notice that the dephasing times 
can become negative (since for instance $\chi_{0}=-{1\over 3}$ 
for $J_{e} = J ={1\over2}$). Although this does not change the sign 
of the 
contribution to the coherent backscattering cone, it might change the 
sign of the correction to the bulk diffusion constant $D$ 
\cite{akma},  
signalling an antilocalization 
contribution like for the spin-orbit correction to the conductivity 
in 
metals. This may lead to an unusual behavior at the Anderson 
localization transition edge. This is quite similar to the 
case of spin-orbit \cite{bergman} in disordered metals (symplectic limit) 
except for the fact that here the problem is richer so that the 
critical exponents depend on the 
Zeeman degeneracy  
$J$ and $J_{e}$ of the atoms 
\cite{amm}.

This research was supported in 
part by the Israel Academy of Sciences and  by the Fund for Promotion 
of
Research at the Technion
as well as by the PRIMA research group of the CNRS in France. C(h+.A).
M. wish to thank A.~Buchleitner and D.~Delande for
their interest.

\end{document}